\documentclass[useAMS,usenatbib]{mn2e}
\usepackage[dv ips]{graphicx,epsfig}

\begin{document}
\title{Extending the Planetary Mass Function to Earth Mass by Microlensing at Moderately High Magnification}
\author[Fumio Abe et al.]{Fumio Abe$^{1}$, Charlotte Airey$^{2}$, Ellen Barnard$^{2}$, Julie Baudry$^{3}$, Christine Botzler$^{2}$, \newauthor Dimitri Douchin$^{4}$,  Matthew Freeman$^{2}$, Patricia Larsen$^{2}$, Anna Niemiec$^{5}$,  Yvette Perrott$^{6}$, \newauthor Lydia Philpott$^{7}$,  Nicholas Rattenbury$^{2}$  and Philip Yock$^{2}$\\ 
$^{1}$Solar-Terrestrial Environment Laboratory, Nagoya University, Nagoya 464-8601, Japan\\
$^{2}$Department of Physics, University of Auckland, Private Bag 92019, Auckland, New Zealand\\ 
$^{3}$D$\acute{e}$partement de Physique, Universit$\acute{e}$ Paris Sud, B$\hat{a}$t 333, F-91403 Orsay Cedex, France\\
$^{4}$Department of Astrophysics, MacQuarie University, NSW 2109, Australia\\
$^{5}$D$\acute{e}$partement de Physique, Ecole Normale Superieue de Cachon, 61, avenue du Pr$\acute{e}$sident-Wilson, 94235 Cachan Cedex, France\\
$^{6}$Astrophysics Group, Cavendish Laboratory, JJ Thomson Avenue, Cambridge, CB3 OHE\\
$^{7}$Department of Earth and Ocean Sciences, University of British Columbia, Vancouver, BC, V6T 1Z4 Canada}
\date{Accepted 2012 August 00. Received 2012 June 00}
\pubyear{2012} \volume{000} \pagerange{1} 
\maketitle \label{firstpage}
\begin{abstract}
A measurement by microlensing of the planetary mass function of planets with masses ranging from $5M_{\oplus}$ to $10M_{\rm{J}}$ and orbital radii from 0.5 to 10 AU was reported recently. A strategy for extending the mass range down to $(1-3)M_{\oplus}$ is proposed here. This entails monitoring the peaks of a few tens of microlensing events with moderately high magnifications with 1-2m class telescopes. Planets of a few Earth masses are found to produce deviations of $\sim5\%$ to the peaks of microlensing light curves with durations $\sim(0.7-3)$ hr in events with magnification $\sim100$ if the projected separation of the planet lies in the annular region $(0.85-1.2)r_{\rm{E}}$. Similar deviations are produced by Earth mass planets in the annular region $(0.95-1.05)r_{\rm{E}}$. It is possible that sub-Earths could be detected very close to the Einstein ring if they are sufficiently abundant, and also planetary systems with more than one low mass planet.    
\end{abstract}


\section{Introduction}
Recently, Cassan et al. (2012) conducted a fairly comprehensive census of planets in the Milky Way using the microlensing technique. For planets at orbital radii of 0.5 - 10 AU they reported that $17^{+6}_{-9}\%$ of stars host planets with masses in the range $(0.3-10)M_{\rm{J}}$, $52^{+22}_{-29}\%$ host cool Neptunes with masses $(10-30)M_{\oplus}$, and $62^{+35}_{-37}\%$ host super-Earths with masses $(5-10)M_{\oplus}$. The results are encouraging for a discipline so young, but the uncertainties are large for the lower mass planets, and also the measurements do not extend down to planets of Earth-mass. In this paper we attempt to address these problems by proposing a modified strategy for detecting low-mass planets by microlensing. 

Our strategy is based on a proposal originally made in 1998 by Griest \& Safizadeh for detecting planets in microlensing events of high magnification. High magnification occurs when the source star passes almost directly behind the lens star, so that $\theta_{\min}\ll\theta_{\rm{E}}$, or $u_{\min}{\ll}1$\footnote{Our notation is fairly conventional. In particular, $r_{\rm{E}}$, $t_{\rm{E}}$ and $t_{0}$ denote the Einstein radius, crossing time and time of closest approach between the lens and source stars, and $\theta_{\rm{E}}$, $\theta_{\rm{min}}$ and $\theta_{\rm{s}}$ the Einstein radius, the impact parameter between the lens and source stars in angular coordinates, and the angular radius of the source star respectively. We denote by $u_{\rm{min}}$ the projected impact parameter in the lens plane between the lens and source stars in units of $r_{\rm{E}}$. Thus $u_{\rm{min}}=\theta_{\rm{min}}/\theta_{\rm{E}}$. Finally, for planetary events, we denote by $q$, $d$ and $\theta$ the planet:star mass ratio, projected separation in units of $r_{\rm{E}}$ and axis relative to the source star track respectively. The fundamental parameter of microlensing, the Einstein radius $r_{\rm{E}}$, is defined in Rattenbury et al. (2002).}. These very well aligned events are rarer than normal microlensing events that have magnifications of only a few, but they are easily recognizable because the peak magnification is approximately $\theta_{\rm{E}}/\theta_{\min}$ and this is $\gg1$. Moreover, as Griest and Safizadeh showed, if a planet orbits the lens star, it almost certainly produces a perturbation to the light curve close to the peak of the light curve. It thus appeared possible to search for planets with high efficiency merely by monitoring the peaks of events that had high magnification.

This was noted by several groups (Gaudi et al. 1998; Rhie et al. 2000; Bond et al. 2002; Rattenbury et al. 2002) and subsequently demonstrated to be correct. Since 2002 a steadily growing number of high magnification events was discovered and alerted each year by the MOA and OGLE collaborations\footnote{MOA: www.phys.canterbury.ac.nz/moa\\OGLE: ogle.astrouw.edu.pl}. Several of these events were monitored by `follow-up' networks\footnote{MicroFUN: www.astronomy.ohio-state.edu/$\sim$microfun\\PLANET: planet.iap.fr\\RoboNet-II: robonet.lcogt.net\\MiNDSTEp: www.mindstep-science.org} and this led to the discoveries of several planets (Udalski et al. 2005; Gould et al. 2006; Gaudi et al. 2008; Bennett et al. 2008; Dong et al. 2009; Janczek et al. 2010; Miyake et al. 2011; Bachelet et al. 2012; Yee et al. 2012; Han et al. 2013), some of the first measurements of planetary abundances (Gould et al. 2006, 2010), and also to a measurement of limb darkening (Fouque et al, 2010). 

A new class of events was also discovered by the MOA and OGLE microlensing collaborations in their quests for high magnifications events. Both collaborations found events having very high magnifications, of order 1,000 (e.g., Abe at al. 2004; Dong et al. 2006). When these were first discovered it was natural to assume they would provide greater sensitivity to planets. This assumption is the subject of the present paper. 

Our conclusion is that low-mass planets can be efficiently searched for in events with moderately high magnifications of order 50-200, and that the greater frequency of these events in comparison to those with higher magnifications lends advantage to monitoring them. However, to take full advantage of their discovery potential, telescopes with apertures in the range 1-2m would be needed to photometer them with sufficient precision to detect low-mass planets\footnote{We note that Los Cumbres Observatory Global Telescope (www.lcogt.net) currently operates telescopes in this aperture range, and has plans for the installation of further telescopes (Brown, T. et al. 2012), and that observations will commence shortly with the Harlingten 1.3m Telescope in Tasmania (John Greenhill, private communication).}.    

Using such telescopes the planetary mass function of Cassan et al. (2012) may be able to be extended down to Earth mass. Because such planets have special importance in the search for the proverbial `Earth twin', and because their abundance will constrain theories of planetary formation, it is important to optimise strategies to encompass these planets in observing programmes. 

Recently Han (2009) and Han and Kim (2009) examined the sensitivity of the high magnification technique for detecting planets. However, they did not explicitly examine the sensitivity as a function of magnification, nor did they focus on the lower magnification events considered in this paper. Our results are therefore complementary to theirs.  
     
\section{Einstein arcs}

Liebes (1964) first suggested that planets might be detected by gravitational microlensing. His analysis was based upon the use of Einstein arcs. Fig. 1 shows the well-known geometry of the pair of moving Einstein arcs that are formed in a normal microlensing event (Paczynski 1996). For clarity, the radius of the source star (projected to the lens plane) has been greatly exaggerated. Typically it is of order ${10^{-3}}\times{r_{\rm{E}}}$ and the Einstein arcs are of comparable width. 

\begin{figure}
\centering
\includegraphics[clip=true,trim=4cm 0cm 0cm 0cm,width=10cm]{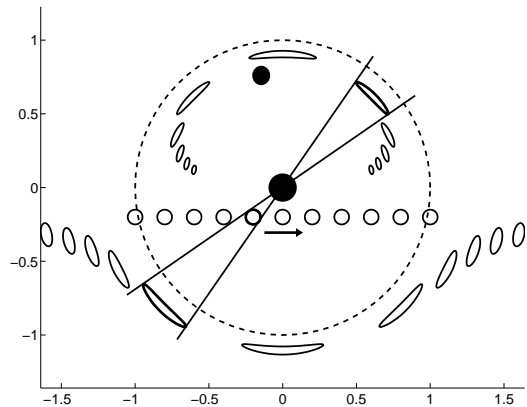}
\caption{The pair of moving Einstein arcs formed in a typical example of gravitational microlensing. The source star is shown as a moving, open circle, and the lens star as a filled, stationary circle. Also shown is a planet in the lens star system that is fairly close to the Einstein ring. The units are fractions of $r_{\rm{E}}$ and the source star radius is scaled to its projected value in the lens plane. The magnitude of $r_{\rm{E}}$ is approximately 2 AU in typical Galactic microlensing events. It is therefore reasonable to suppose the presence of a planet close to the ring as shown in the figure. This figure (without the planet) is adapted from Paczynski (1996).}
\label{f1}
\end{figure}  

Fig. 1 may be used to understand qualitatively the main features of planet detection in high magnification events. Straightforward geometrical considerations show that the lengths of the Einstein arcs in the lens plane near the time of maximum magnification are approximately $2\theta_{\rm{s}}r_{\rm{E}}/\theta_{\rm{min}}$. If the lens star hosts a planet, and if the planet is situated fairly close to either the upper or the lower portions of the Einstein ring as shown in the figure, then one of the Einstein arcs will slide by it during the event, and be perturbed by it. In this way a planet betrays its presence. The effect is large because the image of the source star is magnified by the lens star, and the magnified image is perturbed by the planet. 
 
If the planet is closer to the Einstein ring than the arc length, as shown in the figure, the duration of the planetary perturbation is given by the time required for the arc to slide by the planet. This is easily shown to be approximately $3\rho t_{\rm{E}}$ where $\rho={\theta_{\rm{s}}}/{\theta_{\rm{E}}}$\footnote{Each Einstein arc slides through $120^\circ$ in $t_{\rm{FWHM}}$. Their speeds are therefore ${2\pi r_{\rm{E}}}/{3t_{\rm{FWHM}}}$. This equals ${2\pi r_{\rm{E}} A_{\rm{max}}}/{10.5t_{\rm{E}}}$ (Rattenbury et al. 2002). Dividing the speed into the above expression for an arc length we obtain $3\rho t_{\rm{E}}$ for the slide-by time.}. The result is independent of $u_{\rm{min}}$, and therefore of the peak magnification ${A_{\rm{max}}}$, as $A_{\rm{max}}\approx{u_{\rm{min}}}^{-1}$. Higher magnifications produce longer arcs, but they rotate faster, and the slide-by time is unchanged. The magnitudes of the perturbations are surprisingly large, as discussed below. They are positive if the source star trajectory threads the lens star:planet system ($u_{\rm{min}}\ge0$) and vice versa. As expected, they increase (approximately proportionally) with the mass of the planet, and as the planet approaches the Einstein ring. We note that these scaling laws differ qualitatively from those found for perturbations caused by planetary caustic crossings (Horne et al. 2009).     

Several examples of computed planetary perturbations may be found in the literature that confirm the above expectations (Rhie et al. 2000; Rattenbury et al. 2002; Bennett et al. 2008). For typical values of $\rho\approx0.001$ and $t_{\rm{E}}\approx20$ days planetary pertubations of duration about 1.4 hr are predicted. For typical ranges of $\rho$ and $t_{\rm{E}}$ (see \S6 below) we expect durations ranging over $\sim(0.7-3)$ hr. Such perturbations may be conveniently observed from a single observing site. 

There are two exceptions to the above. If a planet lies very close to the Einstein ring, the difference from the planetless light curve can become large. In this case the above perturbative picture fails, and a quantitative analysis (as carried out below) is required. 

Second, for events with very small impact parameters in which the source star passes behind the lens star, the geometry of the moving arcs is different from that depicted in Fig. 1. In these events a pair of arcs is formed fore and aft of the lens, instead of above and below it (Liebes 1964). A planet perturbs both arcs, one after the other, leading to the formation of two very brief spikes of opposing polarities on the light curve. Examples of such spikes have been observed and modelled (Dong et al. 2009; Janczek et al. 2010; Han \& Kim 2009). Further examples are displayed below. 
   
\section{Magnification maps}

Wambsganss (1997) originally used magnfication maps to demonstrate the detectability of planets by gravitational microlensing. In this method rays are shot from the observer, which is treated as a point source, through the lens in the thin-lens approximation to the source plane to form a `magnification map' that is sampled by the moving source star. 

\begin{figure}
\centering
\includegraphics[clip=true,trim=0cm 0cm 0cm 0cm,width=8cm]{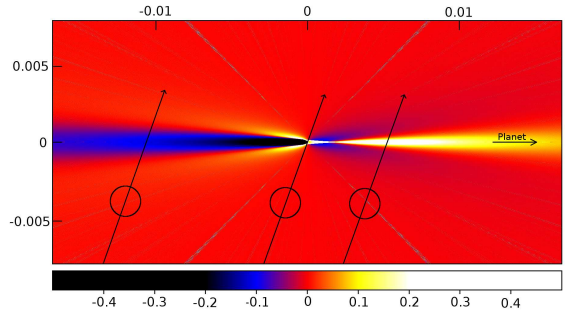}
\caption{Map of normalised fractional planetary deviations produced by a planetary system consisting of a star at the centre and a planet with $q=3\times10^{-5}$ and $d=0.9$ to the right. The deviations were computed pixel-by-pixel as the difference in magnification produced by a star with and without a planet, normalised by the magnification without a planet. The coordinates shown are fractions of $\theta_{\rm{E}}$. Also shown are typical tracks of a solar-like source star at $\theta\approx60^\circ$. Moire-like artefacts caused by the use of rectangular pixels are present. They are alternately positive and negative, and have small net effects when integrated over the source star.}      
\label{f2}
\end{figure}  

Fig. 2 above shows a magnification map for a typical high magnification event. The lens consists of a star at the origin and a planet on the positive $x$-axis at a distance $d$ of $0.9$. The planet:star mass ratio $q$ is $3\times10^{-5}$. This corresponds to a super-Earth of about three Earth masses if the lens star is an M or K dwarf. The map shows the pixel-by-pixel fractional perturbation caused by the planet. It is seen to be approximately $+10\%$ for a large region extending to $x\sim+0.015$ towards the planet, and approximately $-10\%$ for a similarly large region extending to $x\sim-0.015$ away from the planet. 

Also shown are three possible tracks of a source star at a typical inclination to the star:planet axis. The leftmost track is for a solar-like source star with impact parameter $u_{\rm{min}}=-0.01$ or $A_{\rm{max}}=100$. For the central track the impact parameter $u_{\rm{min}}=0$ and $A_{\rm{max}}=2\theta_{\rm{E}}/\theta_{\rm{s}}=2000$ by Liebes's theorem (1964). The righthand track corresponds to $u_{\rm{min}}=+0.005$ and $A_{\rm{max}}=200$. Integrating `by eye' over the area of the source star indicates that the left and right trajectories with $A_{\rm{max}}$ of 100 and 200 yield planetary perturbations approaching $-10\%$ and $+10\%$ respectively in magnitude. 

Perturbations to light curves of order 10\% are readily detectable when the magnification is $\ge100$, yet perturbations of the above type have not been extensively sought in planet searches to date. It should therefore be possible to increase the current detection rate of planets by making use of the large extent of the regions shown on the left and right sides of Fig. 2.        

At the centre of the map lies the `stellar caustic'. This is an approximately triangular closed curve of formally infinite magnification extending from the lens star towards the planet with length about $0.002$, i.e. approximately the diameter of the source star shown in the figure. The true length of the caustic is $0.0027$ (Han 2009) but this is not discernable on the map due to the finite size of the pixels. The caustic provides a means for detecting planets, but source star tracks that traverse it are subject to cancellations from neighbouring areas of opposing polarities. On the other hand, the areas to the left and right of the caustic do not suffer from cancellations to the same extent. In addition, they are longer, and therefore offer the prospect of more numerous traversals in any random sample of events.  

\begin{figure}
\centering
\includegraphics[clip=true,trim=0cm 0cm 0cm 0cm,width=7cm]{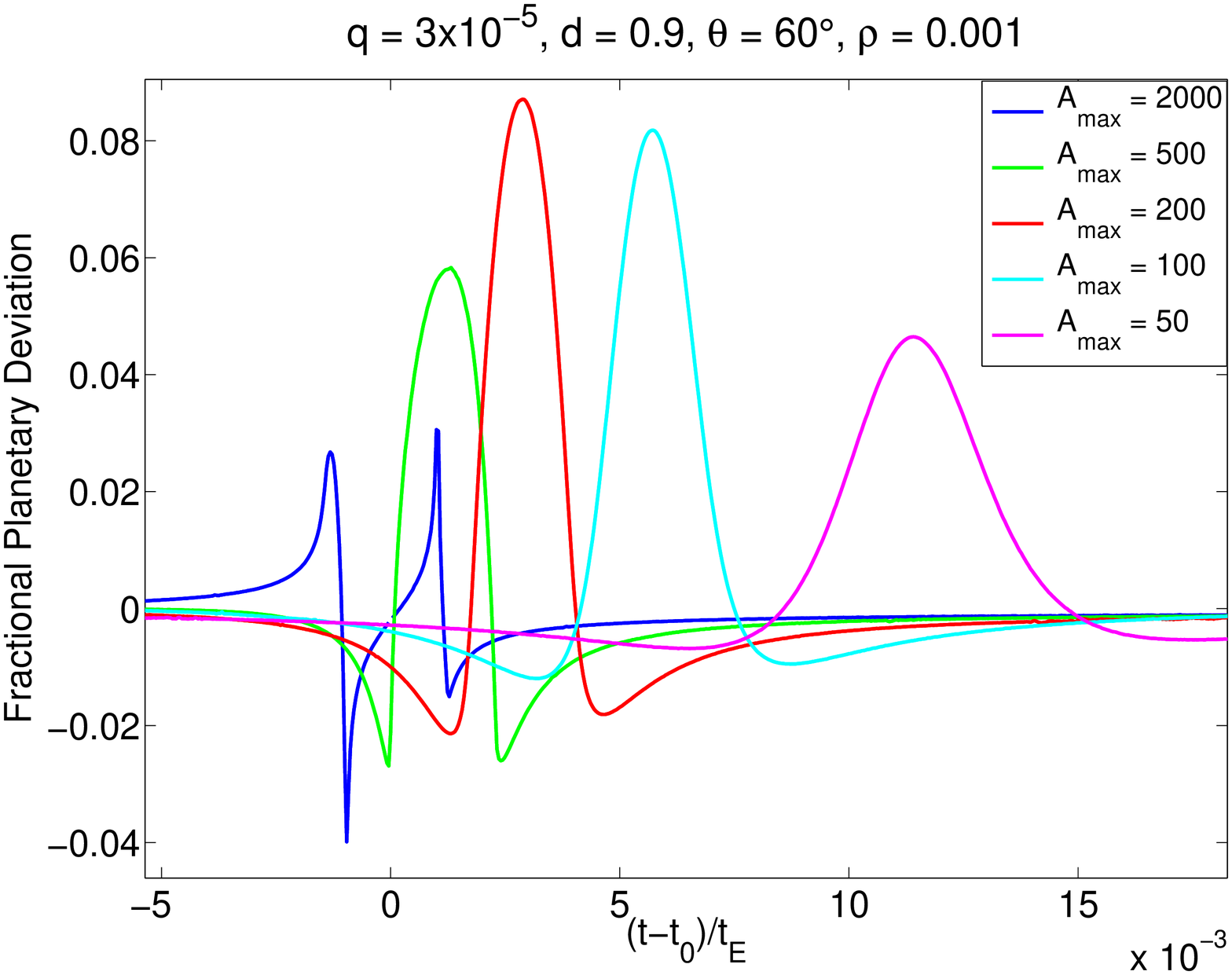}
\caption{Fractional deviations between light curves with and without a planet for various magnifications. The planet:star mass ratio $q$ and separation $d$ are $3\times10^{-5}$ and $0.9$, as in Fig. 2, and the angle between the planet:star axis and the track of the source star is $60^\circ$. The impact parameter $u_{\rm{min}}$ is 0, 0.002, 0.005, 0.01 or 0.02 respectively.} 
\label{f3}
\end{figure} 

Magnification maps similar to that shown in Fig. 2 for heavier or lighter planets, at greater or smaller distances from the Einstein ring, yield similar results. 

\begin{figure*}
\vspace*{5pt}
\includegraphics[clip=true,trim=-1.0cm 0cm 0cm 0cm,width=150mm,height=53mm]{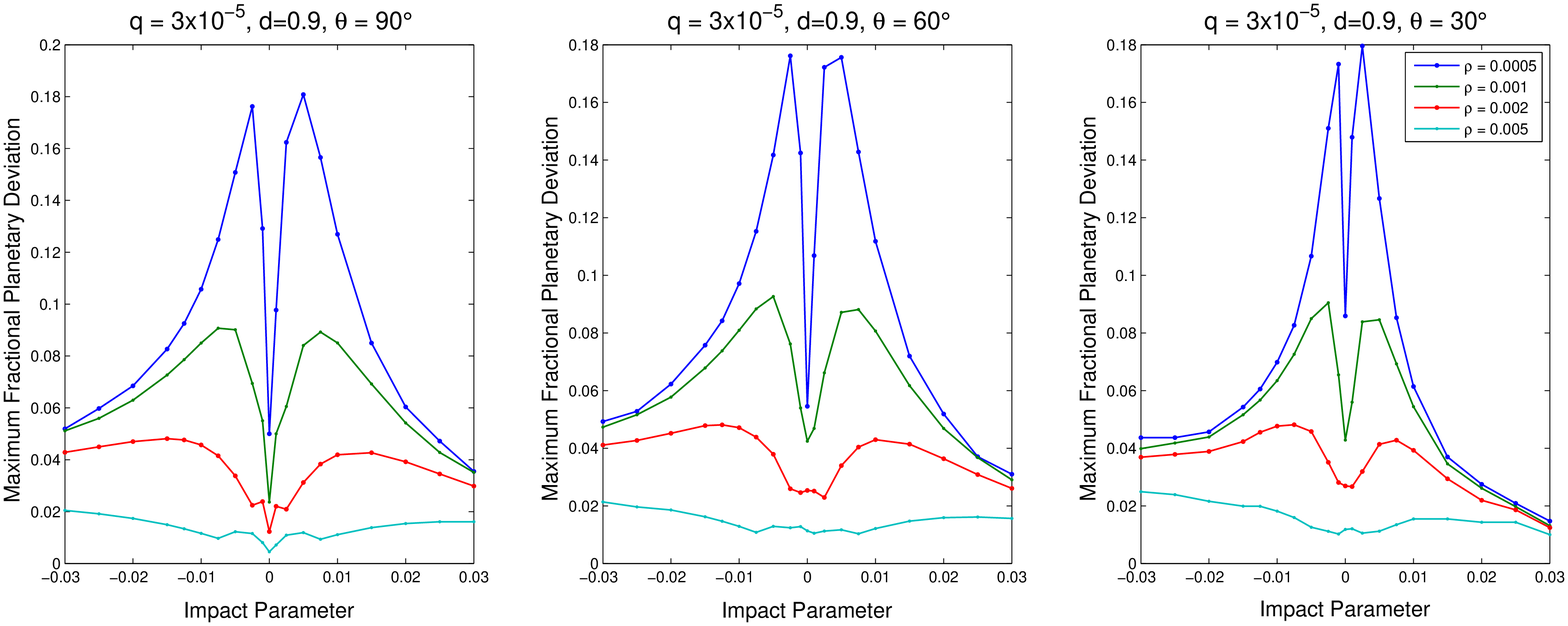}
\caption{Planetary deviations as a function of impact parameter $u_{\rm{min}}$ for various source-star track angles. For each value of $u_{\rm{min}}$ the maximum value of the fractional deviation from the planetless light curve is plotted. Positive values of $u_{\rm{min}}$ correspond to the source star threading the lens system, and vice versa. The planet:star mass ratio and separation are as in Figs. 2 and 3.}
\end{figure*}

\begin{figure*}
\vspace*{8pt}
\includegraphics[clip=true,trim=-1.0cm 0cm 0cm 0cm,width=150mm,height=53mm]{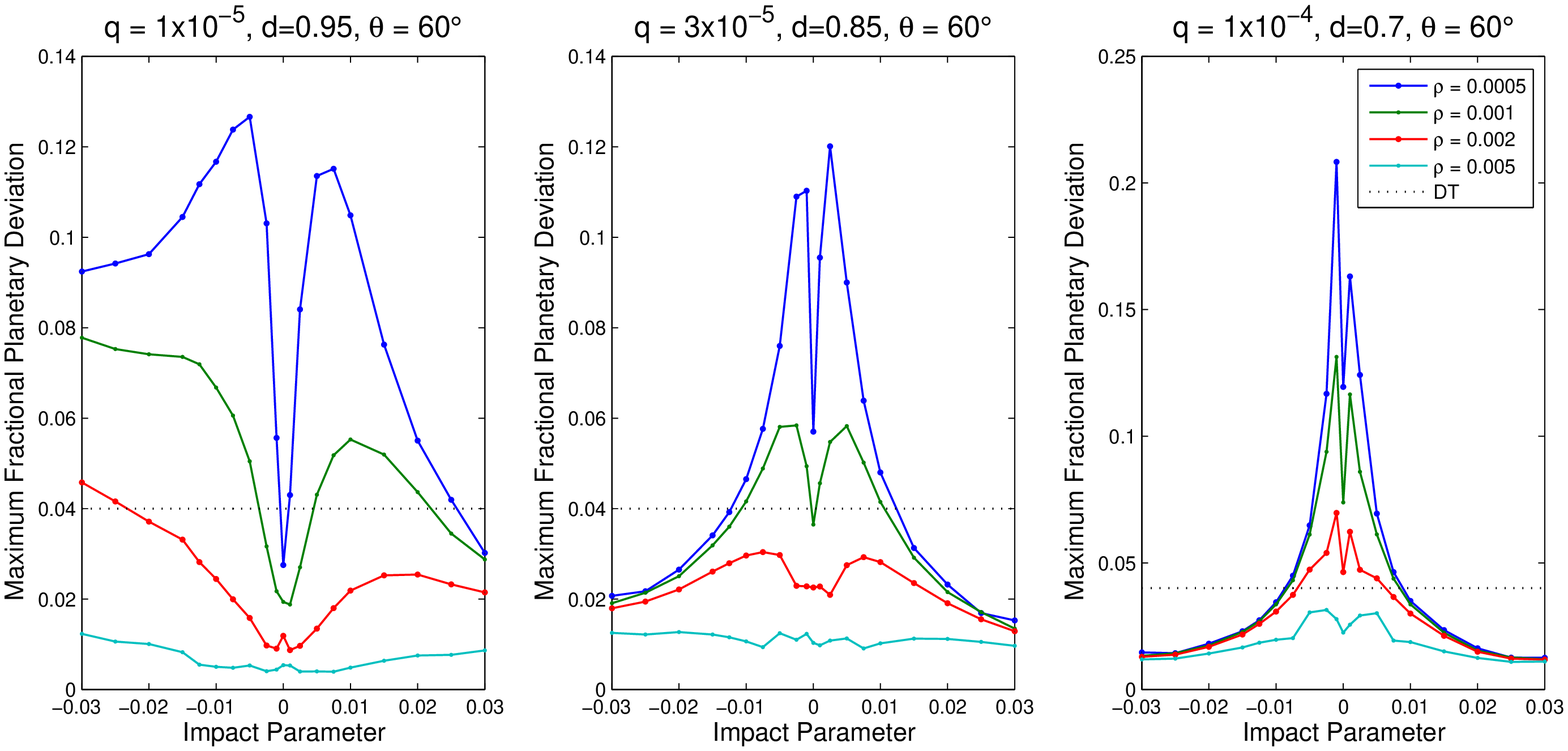}
\caption{Planetary deviations as a function of impact parameter for various planet masses $q$ and separations $d$. A detection threshold of 4\% at $A_{\rm{max}}=100$ was assumed as described in the text. The rising sensitivity at negative $u_{\rm{min}}$ for a planet at $d=0.95$ (and to lesser extents for planets at $d=0.85$ and 0.7) is caused by a cooperative effect between the central and planetary caustics shown in Fig. 6.}  
\vspace*{2pt}
\end{figure*}  

\section{Light curves}
In order to further assess the detectability of low-mass planets, light curves were computed for a range of planet masses and separations. The light curves were generated by subdividing the source plane into three regions. A central region was used to construct the peaks of light curves (approximately $2\times$ the FWHM) at high resolution. A larger region was used to extend the light curves out to approximately $\pm0.5t_{\rm{E}}$ at moderate resolution, and an outer region was used to further extend them to approximately $\pm2t_{\rm{E}}$ at low resolution. The pixel size in each region, and the number of rays per pixel, were adjusted to keep numerical noise below the precision of typical data. The wings of the light curves beyond the outermost of the above regions were treated in the single lens approximation. Limb darkening was included in the linear approximation with a limb darkening coefficient of 0.5. The procedures were checked by comparison with published analyses of several events. 

In order for a planetary deviation to be clearly identified on a light curve, it is helpful if both the deviation and also a significant portion of the unperturbed light curve is observed. If we imagine an ideal search being executed with robotically controlled telescopes programmed to monitor the FWHMs of selected events from sites of excellent seeing, than this requirement will favour events for which $A_{\rm{max}}\le500$. As remarked in $\S2$, the duration of a typical planetary deviation is $\sim3 \rho t_{\rm{E}}$ whereas $t_{\rm{FWHM}}$ is $3.5t_{\rm{E}}/A_{\rm{max}}$ (Rattenbury et al. 2002). To sample both the deviation and the unperturbed light curve equally precisely, the deviation should occupy less than half the FWHM. This translates to $A_{\rm{max}} \le 0.5 \rho^{-1}$ or $A_{\rm{max}}\le 500$ for a typical value of $\rho \approx 0.001.$\footnote{The corresponding condition when hunting exo-moons orbiting free-floating planets can also be met, but not so easily.} We note however that, in actual practice, the descending branch of the FWHM will tend to be more fully monitored than the ascending branch, and that coverage on the descending branch may usefully extend beyond the FWHM. We note also that some previously made planetary discoveries have suffered from less than complete coverage of the FWHM.
        
Fig. 3 shows typical planetary deviations. The planet:star mass ratio and separation are the same as those used in Fig. 2. Fractional deviations are plotted for $A_{\rm{max}}$ equal to 2,000, 500, 200, 100 and 50 respectively. The first exhibits the double-spiked structure discussed above, and the others exhibit perturbations of the width and height predicted above. For the last one at $A_{\rm{max}}=50$ the requirement that the planet be closer to the Einstein ring than the length of the Einstein arcs begins to break down, and the width of the perturbation is greater than predicted.

A number of plots similar to those in Fig. 3 were constructed and examined. Fig. 4 shows the dependence of planetary perturbations on impact parameter $u_{\rm{min}}$. The parameter $u_{\rm{min}}$ was chosen as the independent variable in this plot, rather than the equivalent parameter $A_{\rm{max}}$ which approximately equals $1/{u_{\rm{min}}}$, because $u_{\rm{min}}$ is uniformly populated in any unbiased sample of events, whereas $A_{\rm{max}}$ is not. 

It is apparent that from Fig. 4 that, for all combinations of the other parameters, there is a drop in the fractional deviations for the smallest values of $u_{\rm{min}}$, i.e. for the largest magnifications. This is caused by the onset of the `double-spike' phenomenon discussed above. 

Fig. 4 also shows the dependence of a typical planetary perturbation on the angle $\theta$ between the planet:star axis and the track of the source star. The dependence is essentially constant for angles larger than $30^\circ$. 

\vspace*{10pt}
\begin{figure}
\centering
\includegraphics[clip=true,trim=0cm 0cm 0cm 0cm,width=7.7cm]{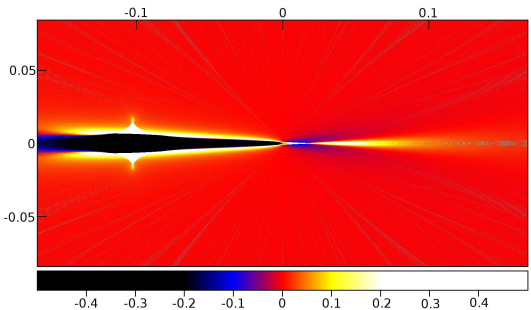}
\caption{Map of fractional planetary deviations for a planet with $q=10^{-4}$ and $d=+0.95$ showing enhanced sensitivity at negative values of $u_{\rm{min}}$ and moderate magnifications.}  
\label{f6}
\end{figure}

The dependence on the source-size parameter $\rho={\theta_{\rm{s}}}/{\theta_{\rm{E}}}$ is also shown in Fig. 4. For typical events with solar-like source stars in the galactic bulge, and main sequence M-K lens stars lying in the bulge or in the galactic disc, $\rho$ takes on values between about 0.0005 and 0.002. It is seen that planetary deviations are greater for smaller values of $\rho$, and they are quite low when $\rho$ is as large as 0.005. The latter value is about the minimum possible value of $\rho$ if the source star is a giant. 

The dependence of typical planetary perturbations on the planet mass parameter $q$ is displayed in Fig. 5. Here $q$ was set equal to $1\times10^{-5}$, $3\times10^{-5}$ or $1\times10^{-4}$. For each value of $q$ the value of $d$ was adjusted (in increments of 0.05) to yield a fractional deviation of approximately 4\% when a solar-like source star with $\rho=0.001$ was magnified 100 times ($u_{\rm{min}}=+0.01$). This procedure yielded approximate ranges of the variables $q,d,u_{\rm{min}}$ and $\rho$ where planets are likely to be detectable. The value of 4\% was chosen as approximately the minimum detectable fractional deviation currently being achieved with 0.3-1.3m class follow-up telescopes. 

The above procedure yielded minimum detectable values of $d$ of approximately 0.95, 0.85 and 0.7 for $q$ equal to $1\times10^{-5}$, $3\times10^{-5}$ and $1\times10^{-4}$ respectively. For values of $d$ lying outside the Einstein ring, the well-known $d$ to $1/d$ symmetry (Griest \& Safizadeh 1998) yields maximum detectable values of approximately 1.05, 1.2 and 1.4 respectively. These limiting values of $d$ agree with those reported by Han (2009) who assumed a minimum detectable fractional deviation of 5\%.

An increase in the breadth of the plots in Fig. 5 as $q$ decreases (right to left) is evident. This implies, irrespective of any details, that very low mass planets with $q\sim10^{-5}$ and $d\sim0.95$ will be detectable in events distributed over a wider range of magnifications than heavier planets with $q\sim10^{-4}$ and $d\sim0.7$. In hunting for the lowest mass planets it will therefore be advantageous to monitor events with the broadest range of magnifications possible. 

We note that the detection efficiences at the above limiting values of $q$ and $d$ are approximately 67\% if only the FWHM's of light curves are monitored intensively. This follows because only those planets with planet:star axes at angles $\ge30^\circ$ to the track of the source star are probed during the FWHM (Rattenbury et al. 2002). 

We note also that the left panel of Fig. 5 exhibits a bias towards negative values of $u_{\rm{min}}$. This is caused by a co-operative effect between the stellar caustic at the origin and a pair of planetary caustics above and below the negative x-axis for a planet located just inside the Einstein ring, as shown in Fig. 6. A similar effect occurs for planets located just outside the Einstein ring and positive values of $u_{\rm{min}}$. 

These cooperative effects already aided discoveries of planets, in MOA-2009-BLG-266 (Muraki et al. 2011) and OGLE-2012-BLG-0461 (C. Han, private communication) for example, and they could boost the detection rate of low mass planets in future events with magnifications of order a few to several tens. However, the lower magnifications of these events may need to be compensated by brighter source stars.

\section{Statistically-adjusted planetary deviations}

The previous discussion can be extended by allowing for the relative precisions with which planetary deviations can be measured at different magnifications. As a first approximation we assume Poisson statistics for the detected photons in any exposure, and multiply fractional deviations by the square root of amplification to approximate relative detectabilities. The results for the three $(q,d)$ combinations shown in Fig. 5 are reproduced below in Figs. 7-9. These include perturbations for negative impact parameters as well as the positive ones shown in Fig. 3. They are almost mirror images of one another (Rattenbury et al. 2002). 

\vspace*{10pt}
\begin{figure}
\centering
\includegraphics[clip=true,trim=0cm 0cm 0cm 0cm,width=7cm]{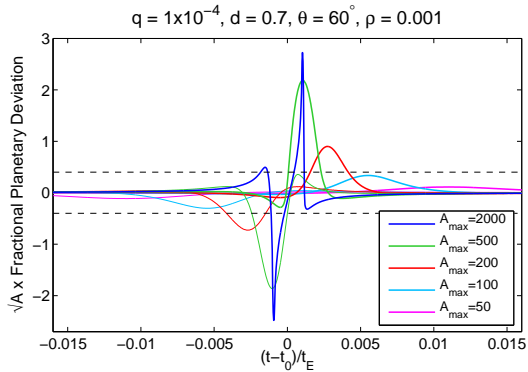}
\caption{Statistically adjusted fractional planetary deviations for $(q,d)=(1\times10^{-4},0.7)$ at various magnifications. The lines at $\pm0.4$ are the estimated limits of detectability with telescopes currently in use as discussed in the text.}  
\label{f7}
\end{figure}

The inclusion of the Poisson factor tends to reverse the effect of the diminished deviations shown in Fig. 3 at large magnifications. In Fig. 7 the largest perturbation occurs at $A_{\rm{max}}=500$. Although the peak signals at $A_{\rm{max}}=2000$ are larger in magnitude, their contributions to $\chi^2$ in a full analysis would be limited by the narrowness of the spikes. 

The estimate made in \S4 of a minimum detectable fractional deviation of 0.04 at $A_{{\rm{max}}}=100$ with telescopes currently in use moves to 0.4 on Figs. 7-9. The calculated perturbations at $A_{\rm{max}}=100$ are seen to be almost at the threshold, as expected.  

The results shown in Figs. 8 and 9 for lighter planets closer to the Einstein ring show the same trend that was found in Fig. 5. The importance of lower magnification events grows until at ($q,d)=(1\times10^{-5},0.95)$ events with magnifications from 50-500 are seen to provide approximately equal sensitivities on an event-by-event basis, with a slight preference for negative impact parameters over positive ones. When one allows for the greater frequency of events with lower magnifications, they are seen to provide the greatest discovery potential.        

\vspace*{10pt}
\begin{figure}
\centering
\includegraphics[clip=true,trim=0cm 0cm 0cm 0cm,width=7cm]{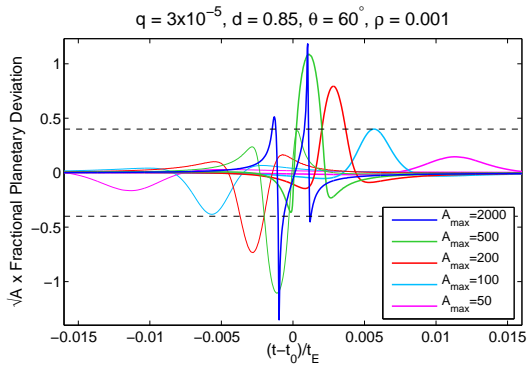}
\caption{Statistically adjusted perturbations for planets with $(q,d)=(3\times10^{-5},0.85)$ at various magnifications. Estimated limits of detectability with current telescopes are shown at $\pm0.4$.} 
\label{f8}
\end{figure}

\vspace*{10pt}
\begin{figure}
\centering
\includegraphics[clip=true,trim=0cm 0cm 0cm 0cm,width=7cm]{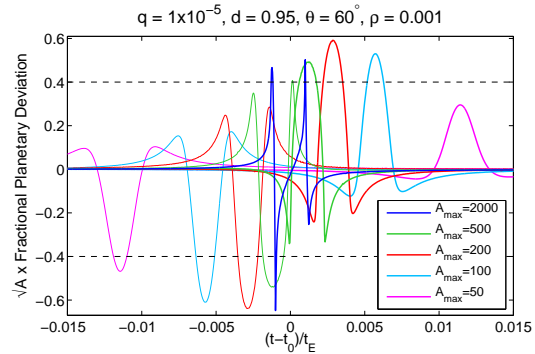}
\caption{Statistically adjusted perturbations for planets with $(q,d)=(1\times10^{-5},0.95)$ at various magnifications. The enhancement at negative impact parameters over positive ones results from the cooperative effect between stellar and planetary caustics discussed in \S4.}   
\label{f9}
\end{figure}

For planets even closer to the ring the advantage of working at lower magnification becomes more evident. Fig. 10 shows results for $(q,d)=(3\times10^{-6},0.98)$. Here it is seen that the sensitivity to sub-Earths is significant at $A_{\rm{max}}$ as low as 50.

\vspace*{10pt}
\begin{figure}
\centering
\includegraphics[clip=true,trim=0cm 0cm 0cm 0cm,width=7cm]{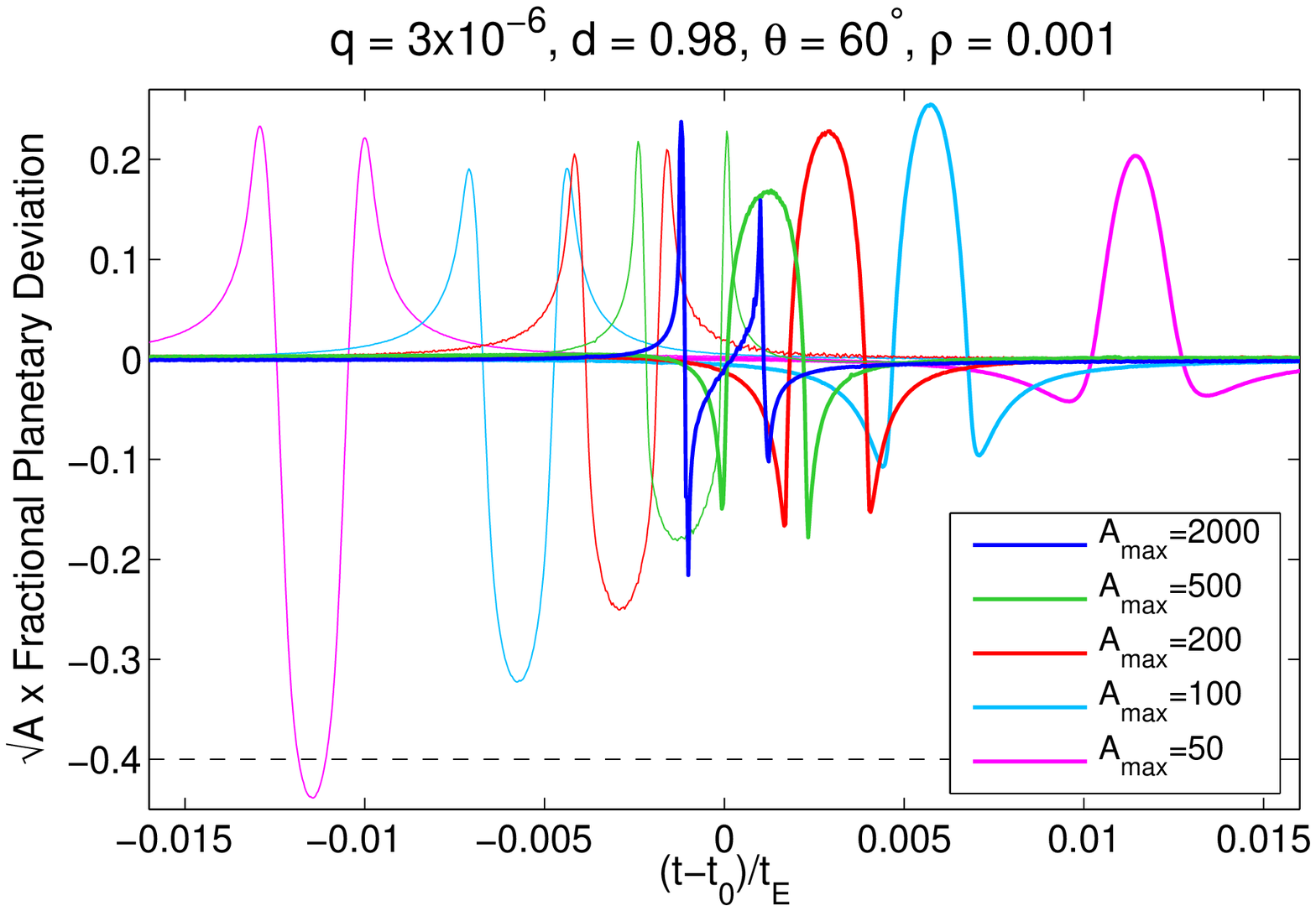}
\caption{Statistically adjusted perturbations for planets with $(q,d)=(3\times10^{-6},0.98)$ at various magnfications. The large enhancement at negative impact parameters results from the cooperative effect between stellar and planetary caustics described in \S4.}   
\label{f10}
\end{figure}    

Fig. 10 exhibits the best of all worlds. The source star is highly magnified, the planet perturbs the highly magnified image, and the magnification is further enhanced by planetary caustics. Moreover, these properties are predictable. Every event with $A_{\rm{max}}\sim50$ will enjoy these advantages and offer the prospect of detecting sub-Earths close to the Einstein ring if they are sufficiently abundant. 
     
\section{Simulation}
To further quantify the detectability of low-mass planets in events of moderately high magnification a sample of simulated events was generated and fully analysed using standard procedures. The simulation was carried out assuming follow-up telescopes with apertures larger than those presently in use become available in the future.     

\subsection{Simulated events}
An expected distribution of events was generated by extrapolating the planetary mass function of Cassan et al. (2002) to planets with masses in the range $(1-6)M_{\oplus}$ and separations in the range 1.25 - 5.0 AU. These ranges were subdivided into a $5\times5$ array as shown in Table 1.  

\begin{table*}
\begin{minipage}{165mm}
\caption{Expected number of planets per 100 stars according to the planetary mass function of Cassan et al. (2012)}
\begin{tabular}{c c c c c c}
\hline\hline
& \multicolumn{4}c{Separation (AU)} \\
& 2.25-2.5 or 2.5-2.78 & 2.0-2.25 or 2.78-3.13 & 1.75-2.0 or 3.12-3.57 & 1.5-1.75 or 3.57-4.17 & 1.25-1.5 or 4.18-5.0\\ 
& \multicolumn{4}c{Projected separation ($d$)}\\
Mass & 0.9-1.0 or 1.0-1.11 & 0.8-0.9 or 1.11-1.25 & 0.7-0.8 or 1.25-1.43 & 0.6-0.7 or 1.43-1.67 & 0.5-0.6 or 1.67-2.0 \\ [0.5ex]
\hline
$(1-2)M_{\oplus}$ & 14.42 & 16.26 & 18.28 & 21.08 & 24.94 \\
$(2-3)M_{\oplus}$ & 5.62 & 6.32 & 7.10 & 8.20 & 9.34 \\
$(3-4)M_{\oplus}$ & 3.08 & 3.48 & 3.90 & 4.50 & 5.34 \\
$(4-5)M_{\oplus}$ & 1.98 & 2.24 & 2.52 & 2.90 & 3.44 \\
$(5-6)M_{\oplus}$ & 1.40 & 1.58 & 1.76 & 2.04 & 2.42 \\ [1ex]
\hline
\end{tabular}
\end{minipage}
\label{t1}
\end{table*}

In the table the relationship of Cassan et al. between the microlensing parameters $q$ and $d$ and physical planet masses and separations was assumed, i.e. $q=1\times10^{-5}$ and $d=1$ correspond to a planet of mass 1 $M_{\oplus}$ at a separation 2.5 AU from its host star. The numbers of planets in the paired ranges in the table are equal according to the mass function of Cassan et al. Also, the efficiencies of planet detection in these ranges are equal according to the $d$ to $1/d$ symmetry of microlensing (Griest \& Safizadeh 1998). These symmetries reduced the number of simulations that were required.  

Magnification maps were constructed for the 25 mass and separation ($q,d$) combinations shown in Table 1. On each map 54 source star tracks were laid down with $\rho$ = 0.0005, 0.001 or 0.002,  $u_{\rm{min}}$ = $\pm0.005$, $\pm0.01$ or $\pm0.02$, and $t_{\rm{E}}$ = 10, 20 or 30 days respectively. The track angle $\theta$ was held fixed at $60^\circ$ because of the approximate independence of detection sensitivity on this parameter shown in Fig. 4 for values of $30^\circ\le\theta\le90^\circ$. It was assumed that these source star tracks would provide a representative sample of real events observed at  moderately high magnification.

Although events with Einstein crossing times $\ge30$ days are common (Sumi et al. 2011), we did not include them in the simulation because less than half of them would enjoy fair weather throughout their FWHM's. Also, we did not weight events with $u_{\rm{min}}$ = $\pm0.02$ more highly than those with $u_{\rm{min}}$ = $\pm0.005$. Even though the former events will occur more frequently, the requirements that the weather be clear throughout the FWHM, and that the FWHM's of consecutive events should not overlap, will favour the latter events.       

The above procedure yielded 1,350 light curves to which noise was added using data from the 1.8m MOA telescope. The MOA telescope was selected for this purpose as it is best known to the authors, and because it is located in a site of less than perfect astronomical seeing. Its performance therefore provides a realistic measure of what is possible with telescopes in the 1-2m class range. 

Published analyses of several MOA events enable the data to be calibrated. The light curve of MOA-2011-BLG-293 yields delta flux values of $\sim7,000$ at I=18 and $\sim17,000$ at I=17 (Yee et al. 2012), MOA-2010-BLG-477 yields a delta flux value $\sim200,000$ at I=14.5 (Bachelet et al. 2012), and MOA-2009-BLG-387 yields delta flux $\sim12,000$ at I=17.4 and $\sim27,000$ at I=16.8 (Batista et al. 2011). These observations that were made on different CCDs on the MOA camera yield an average scale factor of $\sim12,000$ delta flux units at I=17.5 and $\sim120,000$ at I=15. The latter value corresponds to a solar-like star in the Galactic bulge magnified $\sim100\times$ when allowance for typical extinction is included. 

Examination of the above datasets also reveals that the errors at the MOA website\footnote{https://it019909.massey.ac.nz/moa/} are approximately correctly normalized, and that there is a very clear demarkation between data taken in fair weather and data taken in less-than fair weather. For events with main sequence source stars the fair-weather data have photometry errors $\sim300$ delta flux units at all but the highest magnifications. When the delta flux values reach 150,000 - 200,000 the errors gradually rise to 700 - 800. 

Combining the above information we conclude that the precision achieved by the MOA telescope in 60s exposures taken in fair weather when solar-like stars are magnified up to $\sim200\times$ is approximately $0.2\times{A}^{0.2}$ where $A$ denotes magnification. This quite accurately represents the uncertainty when $50\le{A}\le200$ and at lower magnifications. 

Gaussian noise was added to the above sample of light curves using the above parameterization. A sampling rate of one exposure per 0.001 day was assumed for the FWHM of each light curve, reducing to one exposure per 0.02 day out to $\pm3t_{\rm{E}}$, and zero thereafter. The sampling rate of one 60s exposure per 0.001 day is regularly achieved by the MOA telescope, and it could be achieved by follow-up telescopes with smaller cameras. The sampling rate of one exposure per 0.02 day is a realistic estimate of what could occur using the combined resources of the MOA, OGLE, KMTNet and Harlingten telescopes when all are operational. 
    
\subsection{Typical light curves}

The 1,350 simulated light curves were analyzed using a grid minimisation algorithm in which $\chi^2$ was calculated for each combination of parameters on a grid, and the grid was moved until a minimum was found that did not use edge values of any parameter. Initial trials were randomized, and a two-step procedure of coarse steps followed by fine ones was used. The analysis procedure was verified  by comparison with published analyses of several events. 

\begin{figure}
\vspace*{8pt}
\includegraphics[clip=true,trim=5mm 0cm 0cm 0cm,width=90mm,height=65mm]{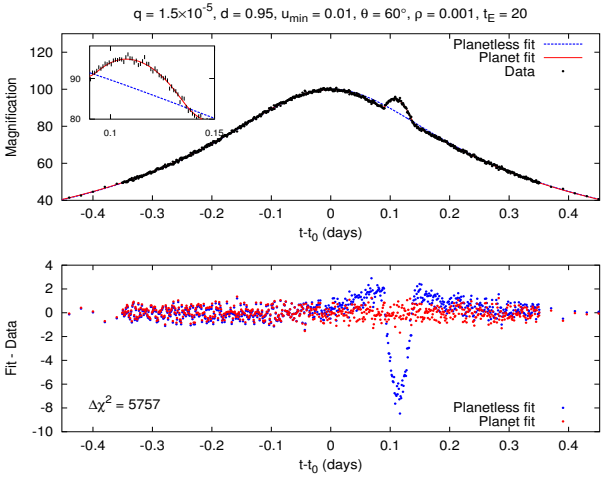}
\caption{Best fits with and without a planet to a simulated light curve from the top left of Table 1, i.e. for the lightest planet in the present study ($q=1.5\times10^{-5}$) at the minimum separation from the Einstein ring ($d=0.95$). The track parameters given in the header led to a clear detection with a $\Delta\chi^2=5757$ difference between the best fits. Most of the 54 tracks laid down on the $(q,d)=(1.5\times10^{-5},0.95)$ map yielded comparably clear detections. The average value of $\Delta\chi^2$ was 9366.}  
\vspace*{2pt}
\label{f11}
\end{figure}

\begin{figure}
\vspace*{8pt}
\includegraphics[clip=true,trim=5mm 0cm 0cm 0cm,width=90mm,height=65mm]{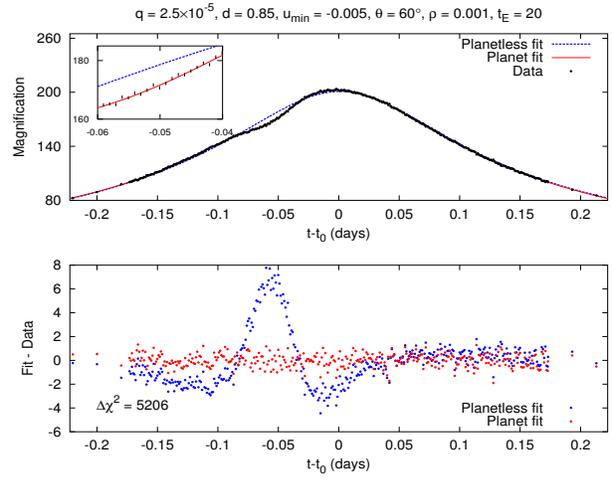}
\caption{Best fits with and without a planet to for a planet with the second lowest mass in Table 1 ($q=2.5\times10^{-5}$) at the second closest position ($d=0.85$) to the Einstein ring for typical track parameters. Many of the 54 tracks laid down on the $(q,d)=(2.5\times10^{-5},0.85)$ map yielded comparably clear detections.}  
\vspace*{2pt}
\label{f12}
\end{figure}  

Typical results for simulated events are shown in Figs. 11 and 12 where low-mass planets were situated fairly close to the Einstein ring. Figs 13 and 14 show extreme examples with the highest mass planet studied here situated closest to the Einstein ring, and a low-mass planet situated far from the ring. The former case leads to a very clear detection ($\Delta\chi^2=323,475$), the latter to no detection ($\Delta\chi^2=4$). Finally, Fig. 15 shows an example of a borderline detection with $\Delta\chi^2\sim300$.
 
\begin{figure}
\vspace*{8pt}
\includegraphics[clip=true,trim=5mm 0cm 0cm 0cm,width=90mm,height=65mm]{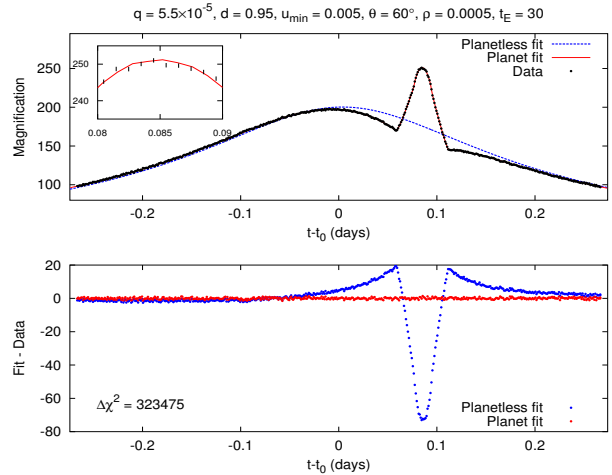}
\caption{Unmistakable signal caused by a `heavy' planet with $q=5.5\times10^{-5}$ close to the Einstein ring ($d=0.95$)}.  
\vspace*{2pt}
\label{f13}
\end{figure}

\begin{figure}
\vspace*{8pt}
\includegraphics[clip=true,trim=5mm 0cm 0cm 0cm,width=90mm,height=65mm]{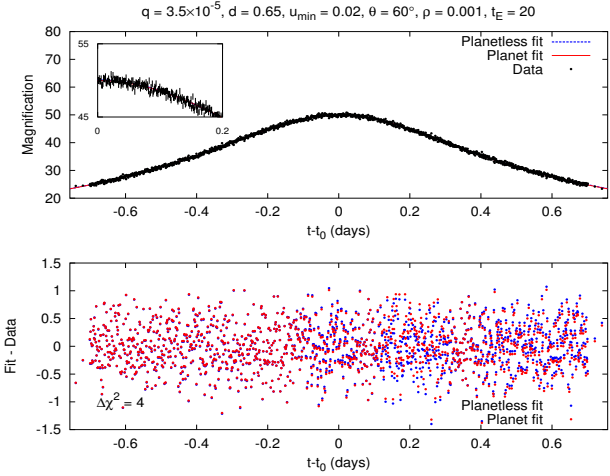}
\caption{Undetectable signal caused by a lightish planet $q=3.5\times10^{-5}$ quite far from the Einstein ring ($d=0.65$)}  
\vspace*{2pt}
\label{f14}
\end{figure}

\begin{figure}
\vspace*{8pt}
\includegraphics[clip=true,trim=5mm 0cm 0cm 0cm,width=90mm,height=65mm]{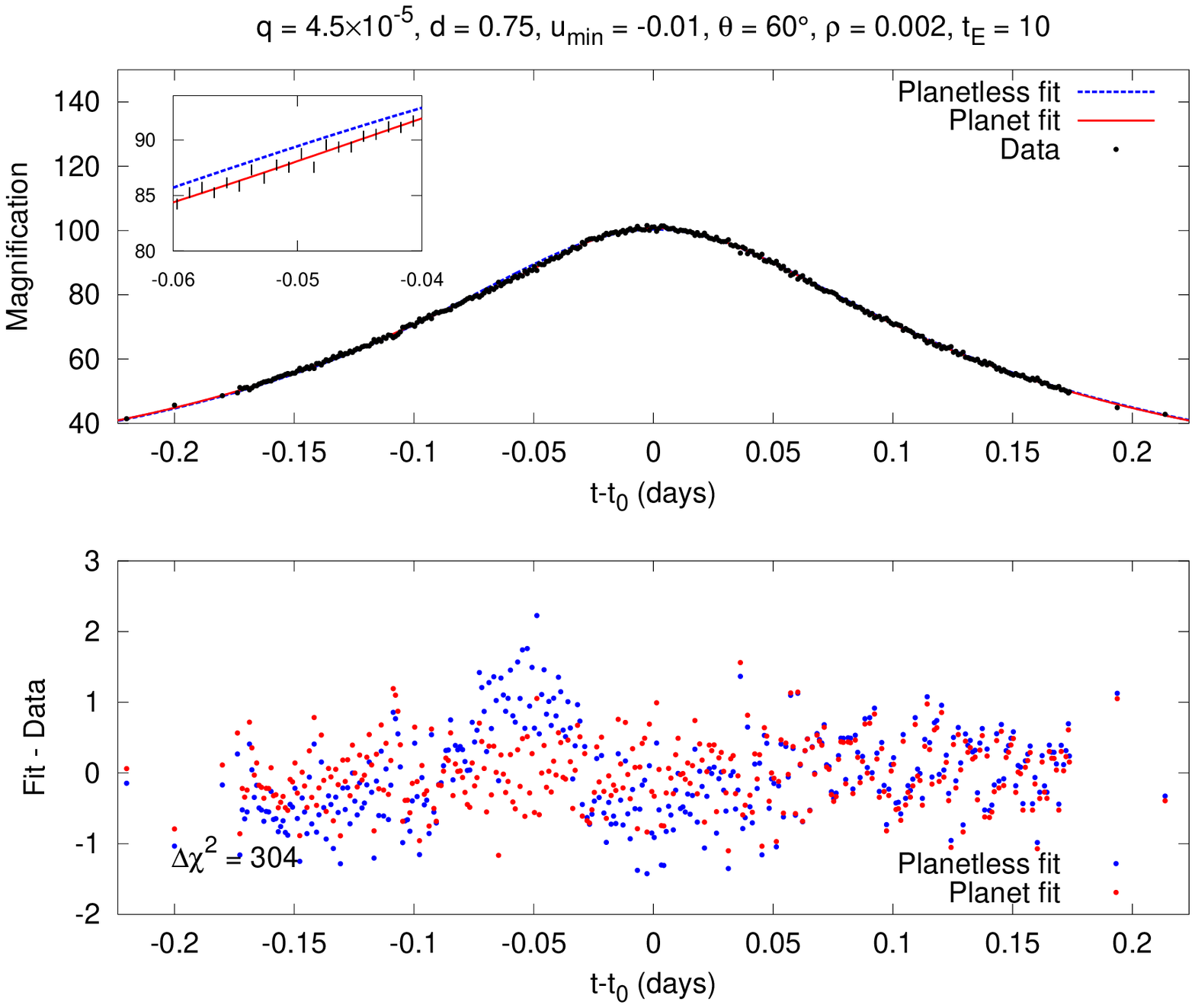}
\caption{Borderline signal with $\Delta\chi^2=304$ caused by slightly heavier planet than that shown in Fig. 13 slightly closer to the Einstein ring. The planetary deviation is approximately 2\%, or only half the detection threshold assumed in Fig. 5.}  
\vspace*{2pt}
\label{f15}
\end{figure}

\subsection{Critical $\Delta\chi^2$ for planet detection}

Several light curves with $\Delta\chi^2\sim300$ were examined by eye to determine a reasonable minimum value of this parameter required for planet detection. Three examples are shown in Fig. 16 with $\Delta\chi^2\le300$ and three with $\Delta\chi^2\ge300$ in Fig. 17. On the basis of these trials we set $\Delta\chi^2=300$ as the minimum value required for detection in our simulation.

This value was chosen for the `ideal' experiment mentioned above, with identical, robotically controlled 1-2m telescopes operating in sites of excellent seeing producing an ensemble of events for joint analysis. It may not be realised in practice. Yee et al. (2013) concluded from an examination of events being monitored with current facilites that $\Delta\chi^2=500$ is a more plausible threshold. 

We note that, if we had chosen the latter threshold in the simulation that follows, the results would not have changed much. Most detections in our simulation have $\Delta\chi^2\ge1000$. In Table 2 that follows, the detection efficiencies for $(1-2)M_{\oplus}$ planets at the first two separations would fall from 0.67 to 0.66 and from 0.36 to 0.26 respectively. Likewise, the efficiencies for $(2-3)M_{\oplus}$ planets at the first two separations would fall from 0.67 to 0.67 (no change) and from 0.53 to 0.44 respectively.    

\begin{figure}
\vspace*{8pt}
\includegraphics[clip=true,trim=5mm 0cm 0cm 0cm,width=90mm,height=65mm]{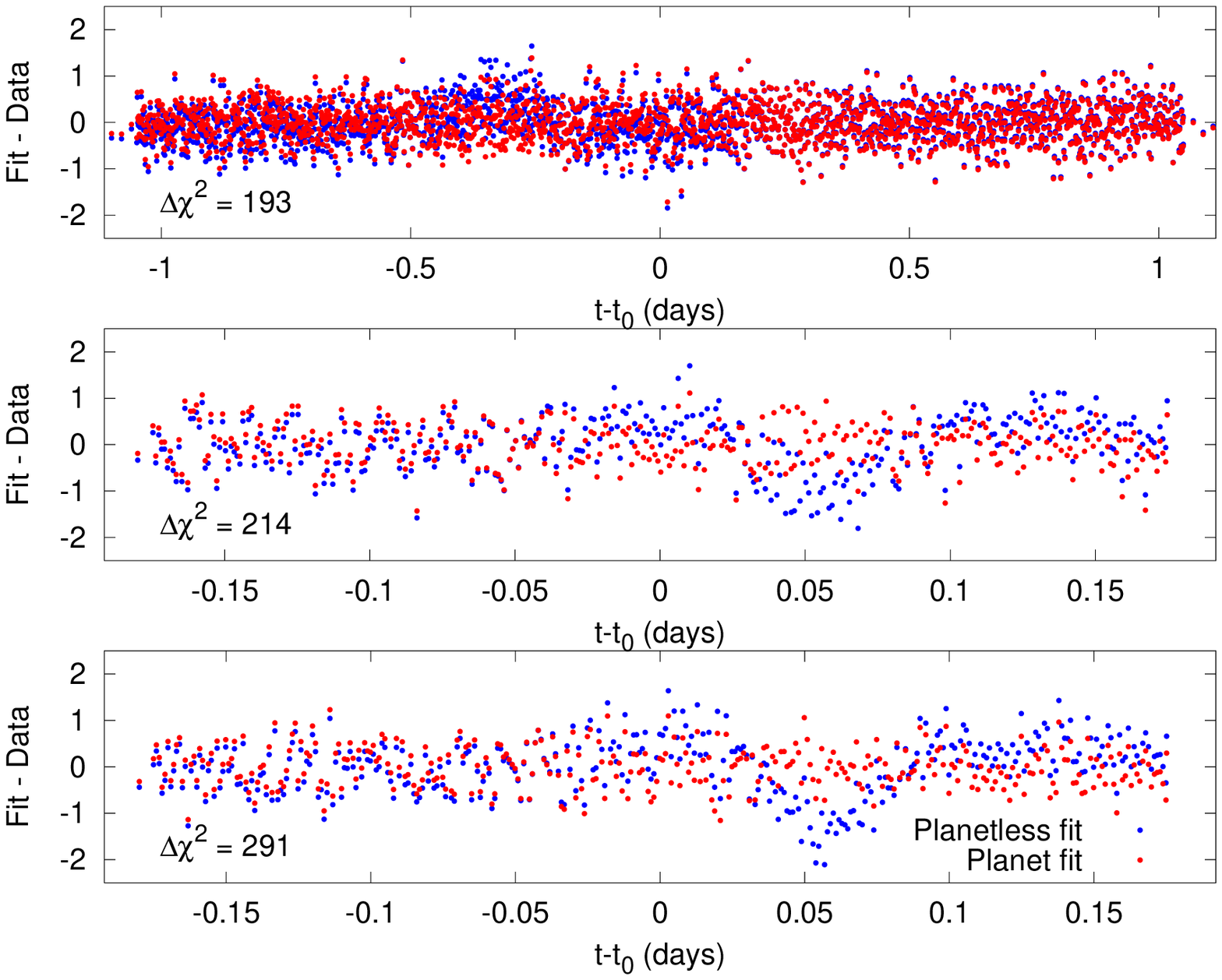}
\caption{Typical planetary deviations with $100\le\Delta\chi^2\le300$.}  
\vspace*{2pt}
\label{f16}
\end{figure}

\begin{figure}
\vspace*{8pt}
\includegraphics[clip=true,trim=5mm 0cm 0cm 0cm,width=90mm,height=65mm]{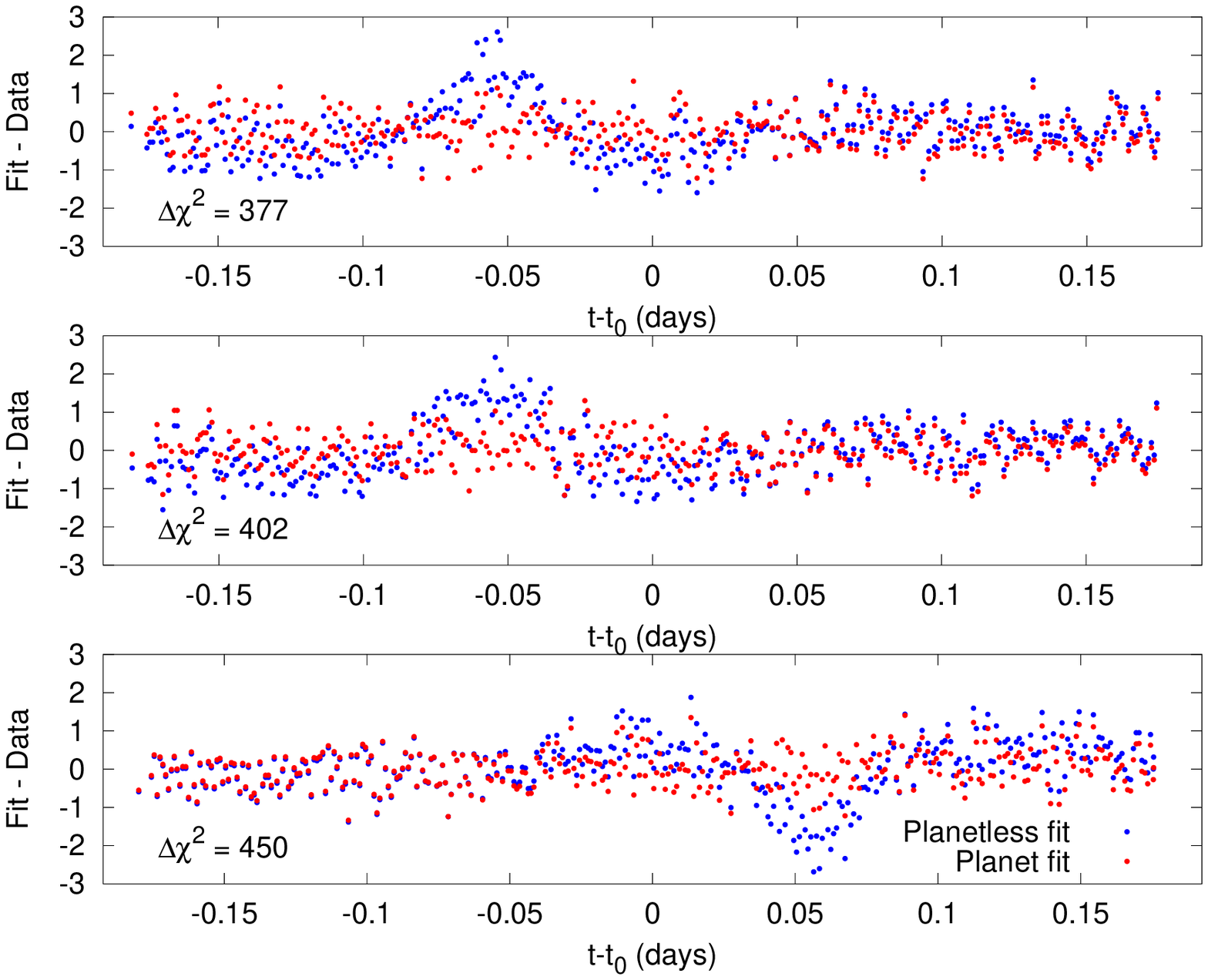}
\caption{Typical planetary deviations with $300\le\Delta\chi^2\le500$.}  
\vspace*{2pt}
\label{f17}
\end{figure}
   
\subsection{Planet detection efficiency}

For each map with $(q,d)$ values given in Table 1 the value of $\Delta\chi^2$ for each of the 54 source star tracks was calculated. An efficiency for planet detection for a given map was determined as $0.67\times$ the fraction of tracks with $\Delta\chi^2\ge300$ where the factor 0.67 was included because only those planets at angles $\ge30^\circ$ are detected during the FWHM (Rattenbury et al., 2002). This yielded the results shown in Table 2. This table also includes the estimated number of detections in each $(q,d)$ interval per 100 stars based on the planetary mass function of Cassan et al. (2012).

It is evident from Table 2 that if some tens of events with moderate magnifications were monitored, and if the planetary mass function of Cassan et al. is approximately valid, then measurements of relatively good statistical precision could be made of the abundance of planets with a few Earth masses.      
  
\begin{figure}
\vspace*{8pt}
\includegraphics[clip=true,trim=-5cm 0cm 0cm 0cm,width=70mm,height=50mm]{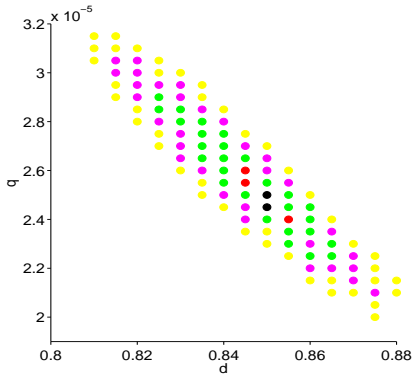}
\caption{$\chi^2$ as a function of $q$ and $d$ for the event shown in Fig. 12 with $\Delta\chi^2=5206$. $1\sigma$, $2\sigma$, $3\sigma$, $4\sigma$ and $5\sigma$ excesses over the minimum are shown in black, red, green, purple and yellow respectively. The known position of the planet in this event, $(q,d)=(2.5\times10^{-5},0.85)$, was well reproduced by the simulated data.}  
\vspace*{2pt}
\label{f18}
\end{figure}

\begin{figure}
\vspace*{8pt}
\includegraphics[clip=true,trim=-6cm 0cm 0cm 0cm,width=70mm,height=50mm]{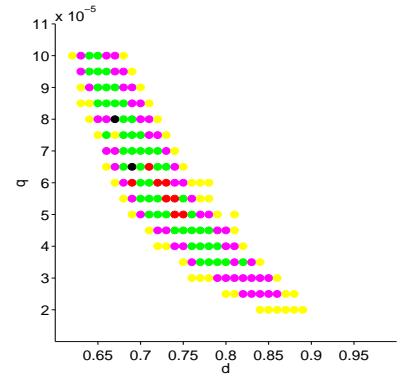}
\caption{$\chi^2$ as a function of $q$ and $d$ for the borderline detection shown in Fig. 15 with $\Delta\chi^2=304$. The colour code is as in Fig. 18. Some numerical noise in the simulated data is evident, and the known position of the planet, $(q,d)=(4.5\times10^{-5},0.75)$, is only approximately reproduced.}   
\vspace*{2pt}
\label{f19}
\end{figure}

\begin{table*}
\begin{minipage}{165mm}
\caption{Efficiency of planet detection in microlensing events of moderately high magnificaion for planets of various masses and separations. The number of detections per 100 stars according to the planetary mass function of Cassan et al (2012) is shown in brackets.}
\begin{tabular}{c c c c c c}
\hline\hline
& \multicolumn{4}c{Separation (AU)} \\
& 2.25-2.5 or 2.5-2.78 & 2.0-2.25 or 2.78-3.13 & 1.75-2.0 or 3.12-3.57 & 1.5-1.75 or 3.57-4.17 & 1.25-1.5 or 4.18-5.0\\ 
& \multicolumn{4}c{Projected separation ($d$)}\\
Mass & 0.9-1.0 or 1.0-1.11 & 0.8-0.9 or 1.11-1.25 & 0.7-0.8 or 1.25-1.43 & 0.6-0.7 or 1.43-1.67 & 0.5-0.6 or 1.67-2.0 \\ [0.5ex]
\hline
$(1-2)M_{\oplus}$ & 0.67 (9.62) & 0.36 (5.82) & 0.11 (2.03) & 0.0 (0.0) & 0.0 (0.0) \\
$(2-3)M_{\oplus}$ & 0.67 (3.75) & 0.53 (3.36) & 0.23 (1.67) & 0.04 (0.30) & 0.0 (0.0) \\
$(3-4)M_{\oplus}$ & 0.67 (2.05) & 0.62 (2.15) & 0.36 (1.40) & 0.12 (0.55) & 0.0 (0.0) \\
$(4-5)M_{\oplus}$ & 0.67 (1.32) & 0.65 (1.46) & 0.42 (1.06) & 0.20 (0.57) & 0.01 (0.04) \\
$(5-6)M_{\oplus}$ & 0.67 (0.94) & 0.67 (1.05) & 0.44 (0.78) & 0.23 (0.48) & 0.06 (0.15) \\ [1ex]
\hline
\end{tabular}
\end{minipage}
\label{t2}
\end{table*}
 
\section{Discussion}
There are a large number of variables that differentiate one gravitational microlensing event from another. It is difficult to allow for every possible contingency even when discussing just one particular characteristic such as planet detection efficiency. In the preceding sections we considered those variables whose values are most likely to affect the detectability of low-mass planets in microlensing events of high magnification. We focused in particular on the peak value of the magnification $A_{\rm{max}}$, but also considered results for various values of the source-size parameter $\rho$ and the Einstein radius crossing time $t_{\rm{E}}$. These variables are the only ones whose magnitudes are likely to be known with any degree of confidence prior to an event reaching peak magnification. On the basis of their values decisions may be made on whether or not to observe an event intensively. 

We have found that events with $50\le {A_{\rm{max}}} \le200$, $0.0005\le{\rho}\le0.002$ and $10\le{t_{\rm{E}}}\le30$ days offer a sensitive hunting ground for low-mass planets provided their FWHM's are monitored continuously with 1-2m class telescopes. The frequency of such events should enable selections to be made to avoid troublesome events such as those with variable or potentially spotty source stars, peak data taken partly in poor weather or under moonlight, or those with nearby bright stars.

\subsection{Accuracy of planetary characterization}
The planetary perturbations shown in the preceding sections are relatively featureless. It might therefore appear difficult to distinguish a perturbation caused by a light planet close to the Einstein ring from that of a heavy planet further away. To quantify this effect we examined the accuracy with which planets could be recovered from their simulated light curves.  

Figure 18 shows $\chi^2 $ as a function of $q$ and $d$ for the light curve shown in Fig. 12 in which $\Delta\chi^2$ was 5,206. In this case the uncertainties in $q$ and $d$ are seen to be $\pm$10\% and $\pm$3\% at the $3\sigma$ level respectively. Fig. 19 shows that the uncertainties grow to $\pm$50\% and $\pm$10\% for the borderline detection in Fig. 15 with $\Delta\chi^2$ was 304. These analyses also yielded uncertainties of $\pm$20\% and $\pm$30\% for the above events at the $3\sigma$ level for the recovered value of the source size parameter $\rho$.   

It is apparent from the above that the richest harvest of planets will occur in events where very low-mass planets are situated close to the Einstein ring. This confirms the expectation given in section $\S4$ that was based on the breadth of the left panel of Fig. 5. It is also evident that definitive detections such as those shown in Figs. 10 and 11 require the use of 1-2m class telescopes.      
\subsection{Terrestrial parallax}
As in other planetary detections by microlensing, measurements of orbital and/or terrestrial parallax can assist to extract absolute values of planetary masses and separations from the microlensing parameters $q$ and $d$. In this regard we note that orbital parallax is unlikely to be measurable in the events we have modelled with $t_{\rm{E}}\le30$ days, but that terrestrial parallax may be measurable in some cases. The rise and fall times of the planetary perturbations shown in Fig. 3 are relatively steep, of order 20 minutes. If these perturbations were intensively monitored from different continents and/or different hemispheres it might be possible to measure terrestrial parallax in some cases. However, we note that terrestrial parallax has been measured previously in events of very high magnification only (Gould et al. 2009; Yee et al. 2009).     

\subsection{Multiplanet events}
If two or more planetary deviations are present on a light curve, as the planetary mass function of Cassan et al. (2012) suggests could be the case when extrapolated to lower masses, it would be easier to identify the deviations if they are well separated. Two conditions must be met to achieve the above. First, the projected position angles of the planets should differ by $\sim30^\circ$ or more in order for the individual deviations not be superimposed on one another, and secondly $t_{\rm{FWHM}}$ must be several times the duration of an individual planetary deviation. This translates to ${A_{\rm{max}}}\le200$ approximately, as shown in Fig. 20.

\vspace*{10pt}
\begin{figure}
\centering
\includegraphics[clip=true,trim=-1.5cm 0cm 0cm 0cm,width=6.5cm]{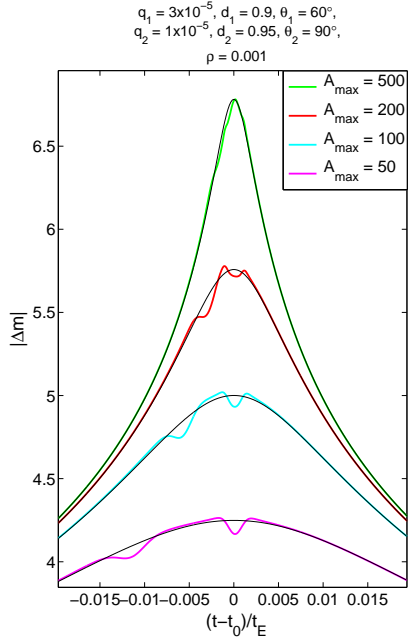}
\caption{Typical planetary deviations in magnitudes for $A_{\rm{max}}$ equal to 50, 100, 200 and 500 for events with two low-mass planets. The planet:star mass ratios are $3\times10^{-5}$ and $10^{-5}$ respectively. Their projected separations $d$ are 0.9 and 0.95 respectively, their position angles differ by $30^{\circ}$, and the impact parameters are negative. Planetary deviations for $A_{\rm{max}}=2000$ would be imperceptible on this plot.}
\label{f20}
\end{figure}
\vspace*{6pt}

We note that multiplanet events have been reported previously (Gaudi et al, 2008; Han et al. 2013) in events with magnifications 100 and 290 respectively. However, the detected planets in these cases were all giants.  
  
\section{Conclusion}

In recent years the microlensing community has focussed attention on events with the highest observed magnifications, in particular on events with magnification $\ge200$ (Gould et al. 2010). As mentioned in \S1 this resulted in the discoveries of several planets by telescopes with apertures ranging from 0.3m to 2.0m. The detected planets have separations from their host stars of order a few AU and correspondingly cool temperatures. Their measured masses range from a few Earth masses to a few Jupiter masses. They provide a representative sample of the distribution of planets between us and the centre of the Galaxy that orbit a relatively unbiased sample of host stars. The discoveries are important because other planetary detection techniques are insensitive to such planets. On the basis of these discoveries the first estimates of the abundances of cool planets in the Milky Way were made.  

In an effort to build on these successes we have proposed a modified procedure for extending the measurements down to Earth mass. The modified strategy makes use of the surprisingly good sensitivity  of microlensing events with magnifications $50\le{A_{\rm{max}}}\le200$ to low-mass planets situated close to the Einstein ring. It was found that high quality detections of low-mass planets such as those depicted in Figs. 11 and 12 could may be made by densely monitoring the peaks of these events with 1-2m class telescopes. If an extrapolation of the planetary mass function of Cassan et al. (2012) to Earth mass is approximately valid, a statistically significant sample of detections could be made by monitoring some tens of events.  

In recent years $\sim10$ events with magnification $\ge200$ have been detected by the MOA and OGLE collaborations annually. The inclusion of events with $A_{\rm{max}}$ down to 50 should quadruple the event rate, and thus permit some tens of these events with moderately high magnifications to be monitored in good conditions relatively rapidly. In this way the planetary mass function could be measured down to Earth-mass. In addition, sub-Earths and planetary systems with more than one low-mass planet might be found. 

Results obtained following the above procedure would provide an independent check of measurements made by monitoring events of low magnification (Kim et al. 2010) and of entirely independent results obtained from radial velocity (Howard et al. 2010; Mayor and Queloz, 2012) and transit measurements (Borucki et al. 2011).

\section*{Acknowledgments}

Support by the Marsden Fund of New Zealand, and comments by Andrew Gould, Gus Hazel, Yasushi Muraki, Ken Prendini, Ian Ramsay, Rachel Street and a referee, are gratefully acknowledged.

\bsp

\label{lastpage}
\end{document}